\documentclass[letter,bibyear]{aa}  
\usepackage{graphicx}
\usepackage{txfonts}

\begin{document}

   \title{The sustained post-outburst brightness of Nova Per 2018,\\ 
          the evolved companion, and the long orbital period\thanks{Table
          1 is only available in electronic form
          at the CDS via anonymous ftp to cdsarc.u-strasbg.fr (130.79.128.5)
          or via http://cdsweb.u-strasbg.fr/cgi-bin/qcat?J/A+A/}}
   \author{U. Munari
          \inst{1}
          \and
          S. Moretti
          \inst{2}
          \and
          A. Maitan
          \inst{2}}
   \institute{National Institute of Astrophysics (INAF), Astronomical
           Observatory of Padova, 36012 Asiago (VI), Italy,\\
              \email{ulisse.munari@inaf.it}
         \and
           ANS Collaboration, c/o Astronomical Observatory, 36012 Asiago (VI), Italy             
             }

   \titlerunning{The evolved companion in V392 Nova Per 2018}   

   \date{Received September 15, 1996; accepted March 16, 1997}

\abstract{Nova Per 2018 (= V392 Per) halted the decline from maximum when it
was 2mag brighter than quiescence and since 2019 has been stable at such a
plateau.  The ejecta have already fully diluted into the interstellar space. 
We obtained {\it BVRIgrizY} photometry and optical spectroscopy of V392~Per
during the plateau phase and compared it with equivalent data gathered prior
to the nova outburst.  We find the companion star (CS) to be a G9 IV/III and
the orbital period to be 3.4118 days, making V392~Per the longest known
period for a classical nova.  The location of V392~Per on the theoretical
isochrones is intermediate between that of classical novae and novae
erupting within symbiotic binaries, in a sense bridging the gap.  The
reddening is derived to be $E_{B-V}$=0.72 and the fitting to isochrones
returns a 3.6 Gyr age for the system and 1.35~M$_\odot$, 5.3~R$_\odot$, and
15~L$_\odot$ for the companion.  The huge Ne overabundance in the ejecta and
the very fast decline from nova maximum both point to a massive white dwarf
(WD) (M$_{WD}$$\geq$1.1$-$1.2~M$_\odot$).  The system is viewed close to
pole-on conditions and the current plateau phase is caused by irradiation of
the CS by the WD still burning at the surface.}
   \maketitle

\section{Introduction}

Prior to its 2018 nova outburst (discovered by Y.  Nakamura on 2018 Apr. 
29.474 UT; cf.  CBET 4515), \object{V392 Per} was a poorly studied cataclysmic
variable (Bruch et al.  1987), eliciting little interest for its rare dwarf
nova outbursts (only a few of which have been recorded) and faintness (mag
17.5 on blue photographic plates during quiescence).  For some time it was
considered lost because of the uncertainty of its position (Bruch et al. 
1987) but was properly identified in later catalogs of cataclysmic variables
(Downes et al.  1997).  Early attempts to record its spectrum were scrubbed
by the faintness of the system (Zwitter \& Munari 1994).  The only published
spectrum of V392 Per prior to its 2018 nova outburst was presented by Liu \&
Hu (2000), who observed the star on 27 Nov 1998, when they estimated it was
at $V$=17.4~mag.

V392 Per attracted wide interest during its recent 2018 nova outburst, in
spite of the quickly approaching conjunction with the Sun.  Spectral
confirmation as a FeII-type nova was soon provided by Wagner et al.  (2018),
who measured FWHM=5200 km\,s$^{-1}$ and E.W.=115 \AA\ for the emission, and
$-$2680 km\,s$^{-1}$ and E.W.=16 \AA\ for the absorption of the P-Cyg
profile affecting H$\alpha$.  Within one day of the outburst announcement, V392~Per
was detected as a strong gamma-ray source by Fermi-LAT (Li et al.  2018). 
Spectra on the following days (Tomov et al.  2018, Darnley et al.  2018a)
supported a widening of the emission profiles (FWHM=5600 km\,s$^{-1}$) and
an even larger velocity for the P-Cyg absorption ($-$4000 km\,s$^{-1}$). 
The main novelty was, however, the splitting of the emission component into
three well separated peaks ($-$2000, $-$250, and $+$1900 km\,s$^{-1}$),
which is a structure destined to last for the rest of the outburst (Darnley 2018a,b;
Munari et al.  2018, 2020; Stoyanov et al.  2020), and a close match to what observed
and modeled by Munari et al.  (2011) in the He/N-type nova V2672 Nova Oph
2009, seen very close to pole-on conditions.

The huge strength they recorded for neon emission lines led Munari et al. 
(2018) to classify V392 Per as a Neon nova, which usually implies a
massive progenitor for the current white dwarf (WD).  Recently, Munari et
al.  (2020) found that V392~Per did not return to quiescence brightness, but
instead remained stuck at $\sim$2 mag above it, reminiscent of novae burning
nuclearly at the surface for years after the end of the outburst (e.g., V723
Cas, Ochner et al.  2015).  We will refer to such "sustained post-outburst
brightness" of V392 Per as SPOB in the rest of this paper.

Radio observations during the first two weeks of the outburst were carried
out with VLA and AMI-LA by Lindford et al.  (2018); the observations clearly
detected the nova at 5.0, 7.0, and 15.5 GHz, and the spectral index
supported a synchrotron emission mechanism.  Finally, it was only following
the solar conjunction that the Swift X-ray/UV satellite  could be aimed at V392
Per.  Darnley et al.  (2018b) reported on these observations and on the
uncertainty in modeling them given the low recorded flux level.

No information is currently available concerning the orbital period of V392
Per, and therefore the evolutionary status of the donor star that Darnley
\& Starrfield (2018) suggest it could be evolved. In this paper we
compare and analyze the spectral energy distribution (SED) of V392~Per in
quiescence and during SPOB, derive the physical properties of the donor
star, and compare these properties with the orbital period of the system we
determined.

\section{Observations}

A low resolution spectrum of V392~Per (range 3500$-$8000 \AA, dispersion
2.31 \AA/pix) was obtained on 2020 Apr 22.833 UT with the B\&C
spectrograph + 300 ln/mm grating mounted on the Asiago 1.22m telescope.  It is
presented in Fig.~1.  It updates the spectrum for 2019 Dec 14 described in
Munari et al.  (2020), which was obtained with an identical instrument set-up.

   \begin{figure}
   \centering
   \includegraphics[width=8.8cm]{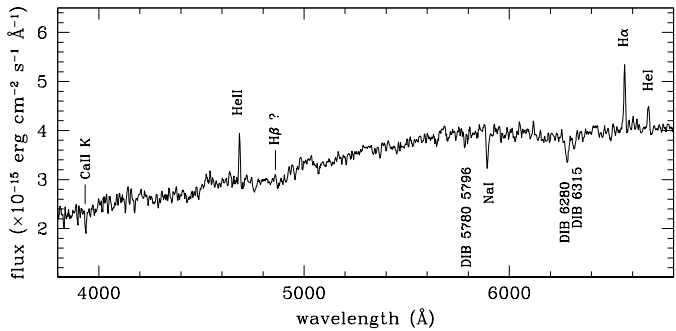}
      \caption{Post-outburst spectrum of V392 Per for 2020 Apr 22.833 UT.}
         \label{f1}
   \end{figure}

Time-resolved optical photometry of V392 Per was obtained in the $V$$R$$I$
bands using the ANS Collaboration 0.42m f/5 Newtonian telescope located in
Bastia (Ravenna, Italy; ANS identifer ID 1205), for the purpose of detecting
orbital modulation; V392~Per was observed for a total of 17 nights and 135
visits, that is to say about eight $V$$R$$I$ set per night on average.  The
presence of a field star of similar magnitude just 8 arcsec to the north
(hereafter FS8) meant the photometry had to be carried out in a PSF-fitting
mode.  The results are given in Table~1.  The quoted error is the
quadratic sum of the Poissonian error on the variable and the error in the
transformation from the instantaneous local photometric system to the
Landolt equatorial standard system.  Additional single-visit photometry,
aimed to map the SED during SPOB, was obtained in Landolt $B$$V$$R$$I$ and
PanSTARRS $g$$r$$i$$z$$Y$ bands with ANS Collaboration ID 310 and Asiago
67/92cm Schmidt telescopes; it is summarized in Table~2.

   \begin{figure}
   \centering
   \includegraphics[width=8.8cm]{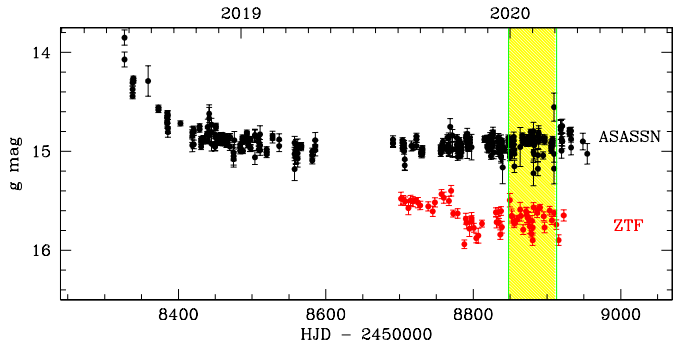}
      \caption{Post-outburst $g$-band lightcurve of V392~Per from ASAS-SN and
      ZTF data.  The region in yellow is the interval covered by our
      time-resolved observations in search of the orbital period.}
         \label{fig2}
   \end{figure}

\begin{table}[t]
\small
\begin{center}
\caption{Time-resolved $V$$R$$I$ photometry of V392 Per (the complete table
is only available in electronic form via CDS (Strasbourg astronomical Data
Center); a portion is shown here to provide guidance on its content).  HJD
is the heliocentric JD - 2458000.}
\label{t1}
\begin{tabular}{ccccccc}
\hline
\multicolumn{7}{c}{}\\
HJD        &   $V$  & err.   &  R     &  err.   &   I    &  err.   \\
\multicolumn{7}{c}{}\\
871.246  & 15.018 & 0.008  & 14.346 & 0.009  & 13.618 & 0.009  \\ 
871.264  & 15.008 & 0.008  & 14.337 & 0.009  & 13.605 & 0.009  \\ 
871.282  & 15.054 & 0.008  & 14.381 & 0.009  & 13.667 & 0.010  \\ 
871.300  & 15.062 & 0.008  & 14.368 & 0.009  & 13.627 & 0.010  \\ 
871.321  & 15.081 & 0.009  & 14.396 & 0.009  & 13.691 & 0.010  \\ 
871.338  & 15.105 & 0.010  & 14.355 & 0.009  & 13.663 & 0.010  \\ 
871.356  & 15.059 & 0.009  & 14.377 & 0.009  & 13.676 & 0.010  \\ 
871.373  & 15.105 & 0.010  & 14.386 & 0.010  & 13.681 & 0.011  \\ 
871.391  & 15.118 & 0.010  & 14.417 & 0.010  & 13.706 & 0.010  \\ 
871.409  & 15.085 & 0.010  & 14.429 & 0.010  & 13.707 & 0.011  \\ 
871.427  & 15.037 & 0.010  & 14.364 & 0.010  & 13.653 & 0.011  \\
\multicolumn{7}{c}{}\\
\hline
\end{tabular} 
\end{center}
\end{table} 

\begin{table}[t]
\small
\begin{center}
\caption{Comparison between photometry prior to the nova outburst
(epoch 2012.838 from PanSTARRS PS1) and during the SPOB phase (2020.194 and
2020.208), used to build the SEDs of Fig. 3.}  
\label{t2}
\begin{tabular}{@{}c@{~~}c@{~~}c@{~~}c@{~~}c@{~~}c@{~~}c@{~~}c@{~~}c@{~~}c@{}}
\hline
\multicolumn{9}{c}{}&\\
$B$&$g$&$V$&$r$&$R$&$i$&$I$&$z$&$Y$& epoch \\
\multicolumn{9}{c}{}&\\
\multicolumn{9}{c}{\underline{\it V392 Per}}&\\
\multicolumn{9}{c}{}&\\ 
        & 17.49   &       & 16.37 &       & 15.74  &          & 15.38  & 15.09 & 2012.838 \\
16.13   & 15.53   & 15.10 & 14.76 &14.46  & 14.28  & 13.73    &        &       & 2020.194 \\
        &         & 14.99 & 14.64 &       & 14.17  & 13.66    & 13.89  & 13.75 & 2020.208 \\
\multicolumn{9}{c}{}&\\
\multicolumn{9}{c}
{\underline{\it field star 8$^{\prime\prime}$ to the north (FS8)}}&\\
\multicolumn{9}{c}{}&\\
        & 15.43   &       & 14.79 &       & 14.50  &          & 14.35  & 14.24 & 2012.838 \\
16.06   & 15.46   &15.08  & 14.82 &14.57  & 14.53  & 14.03    &        &       & 2020.194 \\
        &         &15.09  & 14.81 &       & 14.53  & 14.04    & 14.38  & 14.29 & 2020.208 \\
\multicolumn{9}{c}{}&\\ 
\hline
\end{tabular} 
\end{center}
\end{table} 

\section{Sustained post-outburst brightness (SPOB)}

As first noted by Munari et al.  (2020), in 2019 V392~Per stopped
declining in brightness toward its quiescence value.  This SPOB plateau can
be clearly seen in Fig.~2, which shows the $g$-band lightcurve of V392~Per
reconstructed from observations collected by surveys monitoring the sky for
transients ASAS-SN (Shappee et al.  2014, Kochanek et al.  2017) and ZTF
(Bellm et al.  2019, Masci et al.  2019).  The large difference between the
two sets of data is due to the large pixel-size of ASAS-SN (7.8 arcsec) that
prevents separating V392~Per from FS8.  In Table~2 we compare pre-outburst
photometry of V392~Per (and FS8) from PanSTARRS PS1 (Chambers et al.  2016)
with ours for two SPOB epochs in March 2020.  While FS8 has remained the
same, V392~Per is stuck at 2.0mag brighter than quiescence in $g,$ and this
amount declines smoothly with wavelength down to 1.3mag at $Y$.  This
indicates that the source responsible for the SPOB is hotter than V392~Per
was, as a whole, in quiescence.

The higher temperature conditions of V392~Per during SPOB are confirmed by
the spectrum in Fig.~1 when compared with the quiescence spectrum presented
by Liu \& Hu (2000).  The latter was characterized by
H$\alpha$/H$\beta$$\sim$1 and HeII 4686/H$\beta$$\sim$0.5.  In the spectrum
of Fig.~1, H$\beta$ is possibly too weak to be seen while HeII 4686 is very
strong, rivaling H$\alpha$ in intensity; HeI 6678 also stands out clearly. 
It is also worth noting that the only remaining signature of the nova
ejecta, [OIII] 5007, has further declined compared to its appearance in the
spectrum for 2019 Dec 14, described by Munari et al.  (2020), and is now
just barely visible, marking the final end of the nova phase.

\section{Reddening}

   \begin{figure}
   \centering
   \includegraphics[width=8.8cm]{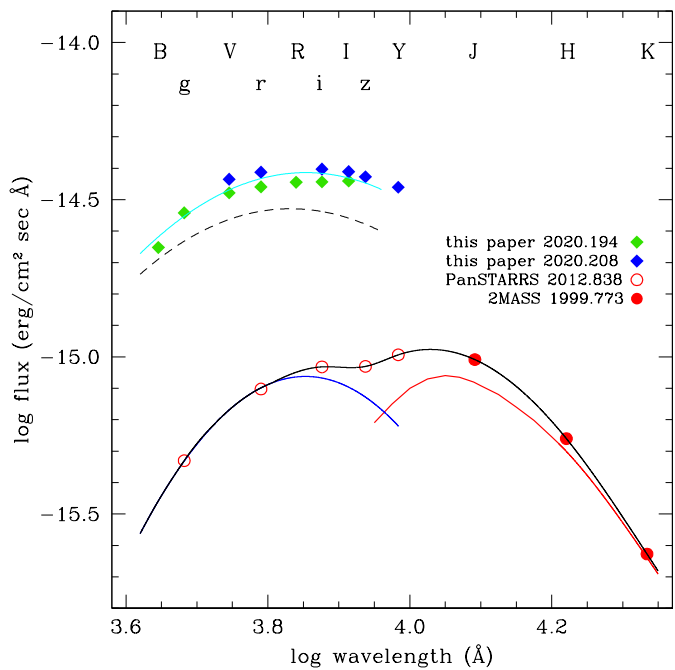}
      \caption{Spectral energy distribution of V392 Per in quiescence (solid
      black line), split into the hot (blue line) and cool (red line)
      components, and during SPOB (cyan line).  
      The dashed black line is the difference between the blue and cyan lines 
      and represents the excess emission during the SPOB phase (see Sect. 4.2).}
         \label{f3}
   \end{figure}

Reddening plays a crucial role in deriving the evolutionary status of the
companion star (CS) to the WD in V392~Per.  There are only two published
estimates, which are in gross disagreement.

The Bayestar2019 3D map of Galactic extinction by Green et al.  (2019)
reports $E_{B-V}$=0.63$^{+0.02}_{-0.03}$ at the Gaia DR2 distance (3.88 kpc)
along the general line of sight to V392~Per.  Tomov et al.  (2018), on their
high resolution ($R_P$=30,000) spectra of V392~Per, found the diffuse
interstellar bands at 5780, 5797, and 6614~\AA, as well as the
interstellar lines of NaI (5890, 5896 \AA) and KI (7665, 7699 \AA) to be
very strong; they derived $E_{B-V}$=1.18$\pm$0.10 from their equivalent
widths.

The high $E_{B-V}$ estimated by Tomov et al.  (2018) is in contrast to the
observed photometric color.  Van den Bergh \& Younger (1987) measured a
mean intrinsic color $(B-V)_\circ$=+0.23$\pm$0.06 for novae at optical
maximum, and $(B-V)_\circ$=$-$0.02$\pm$0.04 at $t_2$.  The AAVSO
lightcurve shows that V392~Per was slightly past maximum when, on 2018 Apr. 
30.116 UT, $B-V$=0.92 was measured by S.  Kiyota (cf.  CBET 4515, where
no estimate of the error is provided), from which we derive 
$E_{B-V}$=0.72$\pm$0.06.

\section{Spectral energy distribution (SED)}

V392~Per in quiescence is present in several all-sky photometric catalogs,
but in most of them it is blended with FS8 and/or the different bands were
observed at widely different epochs.  The only two catalogs resolving
V392~Per from FS8 and presenting simultaneous multi-band data are PanSTARRS
PS1 {\it grizY} bands observed on 2012.194 (Chambers et al.  2016) and 2MASS
{\it JHK$_S$} bands observed on 1999.773 (Cutri et al.  2003).  Their
combined SED is presented in Fig.~3 (black line) and is clearly bi-modal,
with the WD+accretion disk dominating over {\it griz} and the CS on {\it
JHK$_S$}, with the $Y$ band marking the smooth transition between the two. 
This very fact indicates that the brightness of V392~Per was similar at the
time of the PanSTARRS and 2MASS visits, and consequently their data can be
safely combined into a unique SED, marked by the black solid line in Fig.~3. 
We deconvolved it into two components by fitting two second-order
polynomials, represented by the blue and red lines in Fig.~3.  We did not
attempt to fit more complex distributions in view of the very few available
points (eight) constraining the overall fit, and the uncertainty of any
deviation from standard photospheric and disk models due to irradiation and
viewing angle.  The adopted simplification, however, has no practical
consequence on the conclusions reached in this paper.

The original 2MASS photometry of V392~Per in quiescence reads $J$=13.766
($\pm$0.031), $H$=13.290 ($\pm$0.038), and $K_s$=13.062 ($\pm$0.037).  After
subtracting the contribution of the hot component as in Fig. 3, they
become (red line) $J$$\sim$14.099, $H$$\sim$13.392, and $K_s$$\sim$13.092. 
It is relevant to note the very small amount of the subtraction affecting
the $K_s$ band, which constrains the evolutionary status of the CS.

   \begin{figure}
   \centering
   \includegraphics[width=8.8cm]{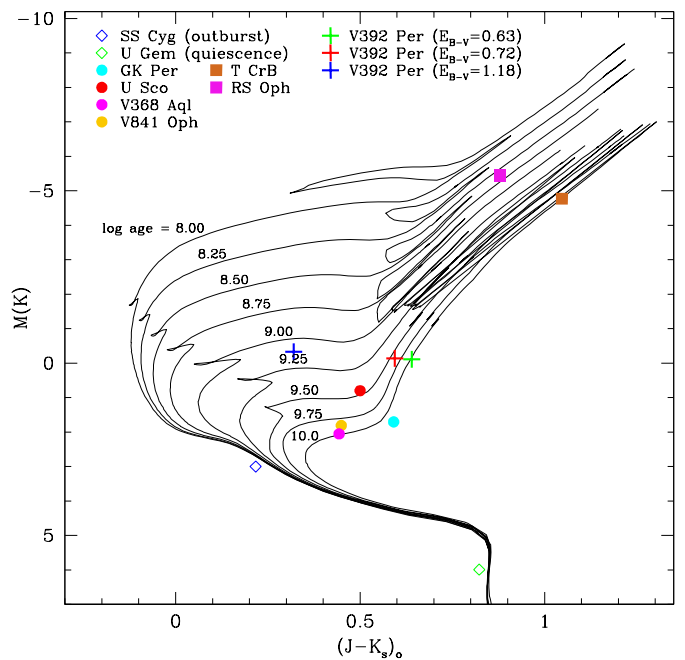}
      \caption{Position of the CS in V392 Per (red line in
      Fig.~3) on the Padova isochrones for the 2MASS plane, dereddened for
      the $E_{B-V}$ values listed in Sect. 4 and Table~4.  Dereddened 2MASS
      photometry of some representative novae and cataclysmic variables with
      accurate Gaia parallax is plotted for comparison (see text for details).}
         \label{f4}
   \end{figure}

\begin{table}[t]
\small
\begin{center}
\caption{Gross properties of the source responsible for extra brightness
of SPOB (the dashed black line in Fig.~3) as
a function of $E_{B-V}$ reddening. The $T_{\rm eff}$ is the temperature
of the fitting black-body and $L_{\rm opt}$ its luminosity
radiated within the optical window (0.3$-$1.0 $\mu$m); $r$ is the radius of
a black-body emitting the equivalent bolometric luminosity.}
\label{t3}
\begin{tabular}{cccc}
\hline
\multicolumn{4}{c}{}\\
$E_{B-V}$  & $T_{\rm eff}$  & $L_{\rm opt}$ & $r$ \\
           & ($^\circ$K)    & (L$_\odot$)   & (R$_\odot$)\\
\multicolumn{4}{c}{}\\
0.63 &  7500 &  42 & 3.85 \\
0.72 &  8750 &  55 & 3.25 \\
1.18 & 38000 & 329 & 0.38 \\
     &       &     &      \\
\hline
\end{tabular} 
\end{center}
\end{table} 

The SED of V392~Per during SPOB has been derived at two nearby epochs,
2020.194 and 2020.208, and plotted in Fig.~3 as green and blue rhombuses
from the data listed in Table~2.  The SED at these two epochs is the same
except for an offset of $\Delta m$$\sim$0.1 mag, which corresponds to the
typical amplitude of brightness variation during SPOB.  A second-order
polynomial fit is plotted as a cyan line.  Subtracting the SED of the hot
component in quiescence (blue line) from it results in the dashed black line
in Fig.~3; this represents the source of the extra brightness of V392~Per during
SPOB.  We have fitted back-body distributions to the dashed line in Fig.~3
as a function of $E_{B-V}$ and the results are reported in Table~3: with
increasing reddening, the temperature and luminosity go up and the
dimensions decline.

\section{Isochrones fitting}

\begin{table}[t]
\small
\begin{center}
\caption{Properties of the CS to the WD in V392 Per from
fitting to Padova isochrones on the 2MASS plane, as a function of $E_{B-V}$
reddening.  The $J$$H$$K_S$ photometry of V392 Per is corrected for the
contribution in quiescence of the hot component (cf. red line in Fig.~3).  The
orbital separation ($a$) and period ($P$) are computed assuming
that CS fills its Roche lobe (see Sect. 6 for details).}
\label{t4}
\begin{tabular}{c@{~~~~}c@{~~~~}c@{~~~~}c@{~~~~}c@{~~~~}c@{~~~~}c@{~~~~}c@{~~~~}c}
\hline
\multicolumn{6}{c}{}\\
$E_{B-V}$ & {\it age}  & $M_{CS}$        & $T^{\rm eff}_{CS}$  & $L_{CS}$      & $R_{CS}$   &  $M_{WD}$   & $a$  & $P$\\
          & (Gyr)  & (M$_\odot$) & (K)    & (L$_\odot$)  & (R$_\odot$)     &  (M$_\odot$)& (R$_\odot$) & (days)\\
\multicolumn{6}{c}{}\\
0.63      & 10.0 & 1.03        & 4740           &    13        &  5.36  & 0.58 & 12.58 & 4.07 \\
          &      &             &                &              &        & 1.10 & 14.49 & 4.38 \\
0.72      &  3.6 & 1.35        & 4875           &    15        &  5.34  & 0.61 & 11.81 & 3.36 \\
          &      &             &                &              &        & 1.10 & 13.37 & 3.61 \\
1.18      &  1.4 & 1.92        & 5915           &    31        &  5.31  & 0.66 & 11.20 & 2.70 \\
          &      &             &                &              &        & 1.10 & 12.39 & 2.90 \\
\multicolumn{6}{c}{}\\ 
\hline
\end{tabular} 
\end{center}
\end{table} 

   \begin{figure}
   \centering
   \includegraphics[width=8.8cm]{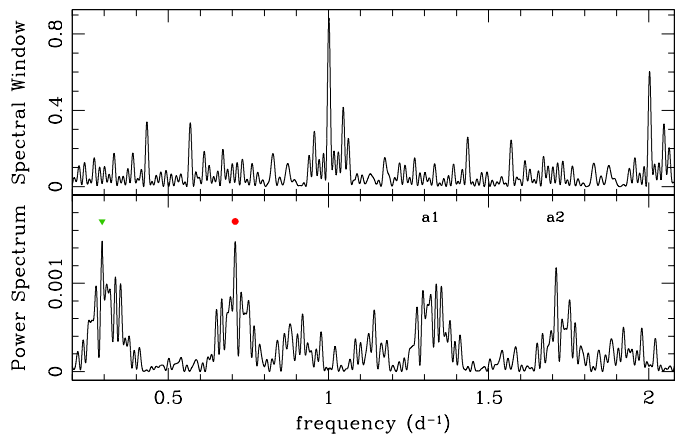}
      \caption{Spectral window and power spectrum over the 5.0$-$0.5 day
               interval for the Fourier analysis of $V$-band data in Table~1.
               The green triangle and red dot mark the 3.4118 and 1.4107 day
               periods, respectively, and $a1$ and $a2$ mark their respective one-day aliases.}
         \label{f5}
   \end{figure}

The 2MASS photometry of the CS in V392~Per (red line in Fig.~3) is compared
with Padova isochrones (Bressan et al.  2012) in Fig.~4 for the values of
$E_{B-V}$ listed in Sect.  4 and Table~3.  We have also plotted the position
of some reference objects: ($a$) two cataclysmic variables of the same UGEM
type as V392~Per, one in quiescence and the other in outburst (U Gem with
Period=4.2 and SS Cyg with P=6.6 hours, respectively); ($b$) four novae that
have some of the longest known orbital periods (P=0.6042 days for
\object{V841 Oph}, P=0.6905 days for \object{V368 Aql}, P=1.2306 days for
\object{U Sco}, and P=1.9968 days for \object{GK Per}); and ($c$) two
recurrent novae with giant secondaries (T CrB with P=227 days and RS Oph
with P=454 days).  The results of fitting to isochrones (for solar
metallicity) is summarized in the first six columns of Table~4.  A higher
reddening corresponds to a younger system and a more massive, more luminous,
and hotter CS.

The remaining columns of Table~4 are devoted to deriving the orbital period
expected under the assumption that the CS (of radius R$_{CS}$) fills its
Roche lobe, as it is customary to assume in cataclysmic variables and novae
(cf.  Warner 1995).  To this aim, two alternatives are considered for the
mass of the WD.  The first is the expected WD mass (from Cummings et al. 
2018) produced by a progenitor, which on the main sequence had a mass 10\%
larger than the current CS.  The second alternative is a WD mass of 1.10
M$_\odot,$ a minimum (Doherty et al.  2015) for a WD of ONeMg composition
taking into consideration the neon overabundance in the ejecta of V392~Per
reported by Munari et al.  (2018).  A high mass for the WD (M$\geq$1.2
M$_\odot$) is also supported by the very short decline times, $t_2$=3 and
$t_3$=11 days, measured by Stoyanov et al.  (2020), relative to Kato \&
Hachisu (1994) theoretical models.

The requirement for the CS to fill its Roche lobe sets the relation between
stellar radius and orbital separation.  If $q$=$M_{CS}$/$M_{WD}$ is the mass
ratio, the orbital separation follows from the Eggleton (1981) relation:
\begin{equation}
\frac{R_{CS}}{a} = \frac{0.49 q^{2/3}}{0.6q^{2/3} + \ln(1 + q^{1/3})}
.\end{equation}
The orbital period in the last column of Table~4 is then computed from
Kepler's third law: ($M_{CS}$+$M_{WD}$)$P^2$=$a^3$.

The main results of Fig.~4 and Table~4 are that the evolutionary status of
the CS requires the orbital period of V392~Per to be longer than those of
classical novae.  The knowledge of the orbital period (see next section)
considerably narrows the uncertainty on properties of the CS left open by
the $\pm$0.06 mag error on the adopted $E_{B-V}$=0.72 reddening.

\section{The orbital period}

   \begin{figure}
   \centering
   \includegraphics[width=8.8cm]{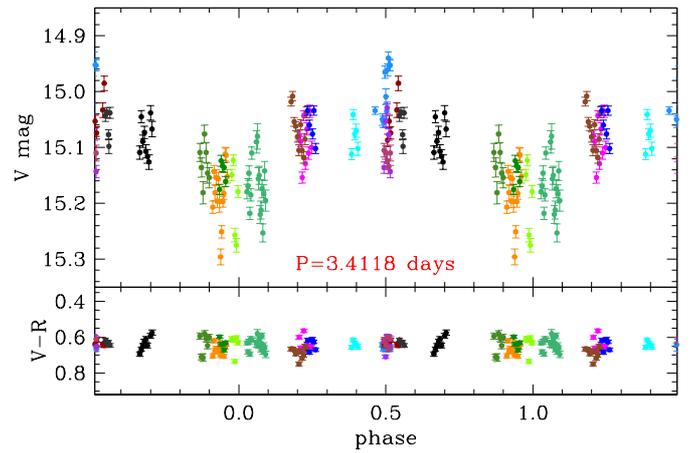}
      \caption{Phased light-curves for the $V$-band data in Table~1
        according to ephemeris in Eq. 2.  Different colors mark data
        from different nights.  The bars are the total error budget listed
        in Table~1.}
         \label{f6}
   \end{figure}

Our time-resolved $V$$R$$I$ observations in Table~1 have been examined for
periodicities.  For the analysis we adopted Fourier transforms following the
formulation by Deeming (1975), and checked the results with the phase
dispersion minimization algorithm of Stellingwerf (1978).  The Fourier power
spectrum and spectral window over the 5.0-0.5 day interval are plotted in
Fig.~5; the results outside this interval are insignificant.

There are two clear peaks in the power spectrum, at 3.4118 and 1.4107 days,
as well as at their one-day aliases (the one-day aliases are marked
respectively with {\it a1} and {\it a2} in Fig.~5).  Their broad pedestals
are caused by the $\sim$70 day duration of the observing campaign and the
grouping of observations to avoid bright Moon phases.  It is worth noting
that the two peaks relate as 1/3.4118 + 1/1.4107 = 1.00 days, so they are
not mutually independent.

Determining which is the true period and which is the alias can be done with the
help of Table~4.  The 3.4118 day period corresponds to highly reasonable
values for the age and mass of the CS, and match the reddening ($E_{B-V}$=0.72)
derived from the photometric color of the nova at maximum.  On the contrary,
a period of 1.4107 days would require an impossibly high value for
$E_{B-V}$ ($\geq$1.8 mag) and a very young age for the system ($\leq$0.3
Gyr), much younger than is typical for cataclysmic variables and novae in
general. The robustness of the 3.4118 day period is also confirmed by
the pre-whitening analysis of the data (see Appendix A). The photometry of
V392~Per in Table~1 is phase-plotted in Fig.~6 following the ephemeris:

\begin{equation}
{\rm min (V)} = 2458853.58(\pm 0.07) + 3.4118(\pm0.0013)\times E
.\end{equation}

\section{System properties}

From 2MASS photometry, orbital period, and the assumption of Roche-lobe
filling, it follows that the CS in V392~Per is an early giant that has just
begun the ascent along the giant branch, and it is characterized by
5.34~R$_\odot$, 1.35~M$_\odot$, 15~L$_\odot,$ and T$_{\rm eff}$=4875~K. 
Such values correspond to a G9~IV-III star (estimated from parameters listed
in Cox 2000).  Such a binary system is highly unusual among classical novae,
and appears to bridge the gap with novae erupting within symbiotic stars, as
its location in Fig.~4 clearly shows.  The presence of an early giant in the
system is in agreement with the very early detection of $\gamma$-rays and
synchrotron radio emission within one day of nova discovery, probably caused
by fast ejecta slamming onto a pre-existing and slow circumstellar medium
formed by the CS wind (cf.  Abdo et al.  2010).  None of the preliminary
descriptions currently available report evidence from spectra of V392~Per of
a deceleration of the ejecta or narrow components from flash-ionized wind
from the companion; however, such deficiencies have already been seen in the
nova outbursts of \object{V3890 Sgr} (Munari \& Walter 2019) and
\object{Nova Sco 2014} (Munari \& Banerjee~2018), which harbor much cooler
and more luminous giants with, presumably, much larger mass-loss rates into
the circumstellar space than V392~Per.

The brightness of V392~Per at the peak of nova outburst is $V$$\sim$6.3 and
the mean in quiescence is $V$$\sim$16.9 (from AAVSO lightcurve).  The
limited $\Delta V$$\sim$10.6 mag outburst amplitude falls much shorter than
the 13.5~mag value implied by the amplitude versus $t_2$ decline-time
relation for face-on systems (Fig.  5.4 in Warner 1995, updated to the
zero-point by Selvelli \& Gilmozzi 2019), with an obviously increasing
discrepancy for larger orbital inclinations.  The brighter-than-usual
magnitude in quiescence is further evidence of the presence of a CS in
V392~Per that is much brighter than in classical novae, whose optical
brightnesses are completely dominated by the accretion disk.  In this sense,
it is worth noting that at the distance (3.88$^{+0.98}_{-0.64}$ kpc, from
Gaia DR2 parallax) and reddening ($E_{B-V}$=0.72) affecting V392~Per, the
G9~IV-III companion would shine at $V$$\sim$17.35, adopting the absolute
magnitude tabulated by Sowell et al.  (2007).  To add to $V$$\sim$16.9 in
quiescence, the WD+disk needs to emit at $V$$\sim$18.0, brighter than the
$V$$\sim$18.8 implied by the MMRD relation (magnitude at maximum versus rate
of decline) as calibrated on Gaia distances by Selvelli \& Gilmozzi (2019),
and scaled to face-on orbital inclination.  The wider orbital separation and
higher mass transfer in V392~Per are able to support a disk that is larger and
brighter than in classical novae.

If a low orbital inclination of V392~Per is favored by the above line of
reasoning on the brightness in quiescence, it is actually required by the
profile of emission lines observed during the 2018 nova outburst.  The bulk
velocity of the approaching and receding components of the bipolar flow
($-$2000 and $+$1900 km~s$^{-1}$; Tomov et al.  2018) are among the largest
ever observed in classical novae; such a wide split can only be obtained for
a low orbital inclination and close to face-on orientation.  At low orbital
inclination, the side of the CS that is irradiated by the burning WD is in
constant sight, and a limited deviation from $i=0^\circ$ is responsible for
the small amplitude ($\Delta m$$\sim$0.1 mag) of the orbital modulation
(Fig.~6).  The constancy of photometric color along the orbital motion
(Fig.~6) agrees with the inferred low inclination.  For comparison, the
orbital modulation caused by irradiation in V723~Cas amounts to $\sim$2 mag
given the much larger orbital inclination of that system ($\sim$65$^\circ$,
Ochner et al.  2015).  The irradiated side of the CS as the source of the
extra optical brightness of V392~Per during SPOB is confirmed by the large
dimension derived in Table~3 for the $E_{B-V}$=0.72 case (implied by the
P=3.4118 day orbital period and confirmed by the color at nova maximum). 
This scenario is also in agreement with the small equivalent width of the
emission lines in the spectrum shown in Fig.~1: they need the high
temperature of an irradiated disk to form but have to emerge over the
rethermalization spectrum of the irradiated side of the CS (which is too
cold to originate any of these emission lines, cf.  Table~3).

\begin{acknowledgements}
We thank P.  Selvelli (INAF Trieste) and R.  Raddi (UC, Barcelona) for
useful discussions and P.  Ochner (Univ.  of Padova), S.  Dallaporta, A. 
Frigo, V.  Andreoli, M.  Graziani and S.  Tomaselli (ANS Collaboration), for
collecting some of the photometric data used in this project.
\end{acknowledgements}

\begin{appendix}
\section{Fourier analysis of pre-whitened photometric data}

A Fourier analysis was carried out on the same input photometric data
used for the periodogram of Fig.~5, after pre-whitening for the 3.4118 day 
period. The results are shown in Fig.~A.1. The absence of residual  
peaks confirms the correctness of the derived period. 

   \begin{figure}[!h]
   \centering
   \includegraphics[width=8.8cm]{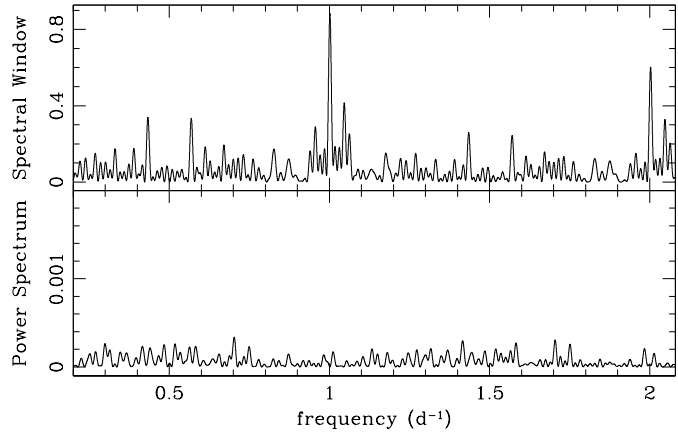}
      \caption{Spectral window and power spectrum over the 5.0$-$0.5 day
               interval for the Fourier analysis of $V$-band data in Table~1
               after pre-whitening for the 3.4118 day period.}
         \label{f8}
   \end{figure}

\end{appendix}
\end{document}